\documentclass{article}
\PassOptionsToPackage{numbers, compress}{natbib}
\usepackage[preprint]{neurips_data_2023}

\usepackage[utf8]{inputenc} 
\usepackage[T1]{fontenc}    
\usepackage{hyperref}       
\usepackage{url}            
\usepackage{booktabs}       
\usepackage{amsfonts}       
\usepackage{nicefrac}       
\usepackage{microtype}      
\usepackage{xcolor}         

\usepackage{amsmath}
\usepackage{graphicx}
\usepackage{comment}
\usepackage{subcaption} 

\usepackage{ulem}
\definecolor{Vert}{RGB}{0,128,0}

\title{Forecasting Electric Vehicle Charging Station Occupancy: Smarter Mobility Data Challenge}

\author{
  Yvenn Amara-Ouali \\ 
  EDF R\&D\\
  \And
  Yannig Goude \\
  EDF R\&D\\
  \And
  Nathan Doumèche\\
  EDF R\&D\\
  \And
  Pascal Veyret \\
  EDF R\&D\\
  \AND 
  Alexis Thomas\\
  Ecole des Mines de Paris\\
  \And
  Daniel Hebenstreit\\
  Graz University of Technology\\
  \And 
  Thomas Wedenig\\
  Graz University of Technology\\
  \And
  Arthur Satouf\\
  CY Tech\\
  \And 
  Aymeric Jan\\
  SLB, AI Lab\\
  \And 
  Yannick Deleuze\\
  Veolia S\&TE
  \And 
  Paul Berhaut\\
  Air Liquide
  \And
  Sébastien Treguer \\
  INRIA
  \And
  Tiphaine Phe-Neau \\
  Renault Digital\\
}

\begin{document}

\maketitle

\begin{abstract}
The transport sector is a major contributor to greenhouse gas emissions in Europe. Shifting to electric vehicles (EVs) powered by a low-carbon energy mix would reduce carbon emissions. However, to support the development of electric mobility, a better understanding of EV charging behaviours and more accurate forecasting models are needed. To fill that gap, the Smarter Mobility Data Challenge has focused on the development of forecasting models to predict EV charging station occupancy. This challenge involved analysing a dataset of 91 charging stations across four geographical areas over seven months in 2020-2021. The forecasts were evaluated at three levels of aggregation (individual stations, areas and global) to capture the inherent hierarchical structure of the data. The results highlight the potential of hierarchical forecasting approaches to accurately predict EV charging station occupancy, providing valuable insights for energy providers and EV users alike. This open dataset addresses many real-world challenges associated with time series, such as missing values, non-stationarity and spatio-temporal correlations. Access to the dataset, code and benchmarks are available at \href{https://gitlab.com/smarter-mobility-data-challenge/tutorials}{https://gitlab.com/smarter-mobility-data-challenge/tutorials} to foster future research.

\end{abstract}

\section{Introduction}

\label{intro}

\paragraph{Electric mobility.}
The transport sector is currently one of the main contributors to greenhouse gas emissions in Europe \cite{EEA_2022}. The development of electric vehicles (EVs) combined with a low-carbon energy mix would make it possible to considerably reduce these emissions. Fortunately, the EV initiative is one of the few programmes listed by the International Energy Agency (IEA) that aligns with the IEA's net-zero emissions goals. In 2021, China led global EV sales with 3.3 million units, tripling its 2020 sales, followed by Europe with 2.3 million units, up from 1.4 million in 2020. The U.S. market share of electric vehicles doubled to 4.5\%, with 630,000 units sold. Meanwhile, electric vehicle sales in emerging markets more than doubled \cite{IEA_EV_2022}. Electric mobility development entails new needs for energy providers and consumers \cite{rte2022}. Companies and researchers are proposing innovative solutions including pricing strategies and smart charging \cite{dallinger2012grid, wang2016smart, alizadeh2017optimal, Moghaddam2018smart, crozier2020a}.  
However, their implementation requires a precise understanding of charging behaviours and better EV charging models are necessary in order to better understand the impact of EVs on the grid \cite{kaya2022electric, ciociola2023data}. In particular, forecasting the occupancy of a charging station can be a critical need for utilities to optimise their production units according to charging demand. On the user side, knowing when and where a charging station will be available is critical, but large-scale datasets on EVs are rare \cite{calearo2021a, amara2021review}. This is why the dataset provided in this challenge is valuable in itself. This challenge aims at testing statistical and machine learning forecasting models to forecast the states of a set of charging stations in the Paris area at different geographical resolutions.

\paragraph{Hierarchical forecasting}
The problem of the challenge presents an intrinsic structure, called hierarchy, since it consists in forecasting quantities at increasing scales (stations, areas, and global). Hierarchical time series forecasting has been studied for various other applications where the data is directly hierarchically organised or where there is a latent hierarchical representation. For example, in retail, goods are often classified into categories (such as food or clothing) and inventory management can be done at different geographical (national, regional, shop)  or temporal (week, month, season) scales. Moreover, electricity systems often have an explicit (electricity network) or implicit (customer types, tariff options...) hierarchy. Recent work shows that exploiting this structure can improve forecasting performance at different levels of hierarchy. For instance, \cite{hyndman2011optimal} focuses on tourism demand,  \cite{Athanasopoulos2019HierarchicalF} on macroeconomic forecasting, and \cite{Hong2019GlobalEF, Brgre2020OnlineHF, Taieb2020HierarchicalPF} on electricity consumption data.

\paragraph{Overview.}

The paper is structured as follows. Section \ref{EV_charging} describes the forecasting problem at hand and the baseline models. Section \ref{winning} presents the methods proposed by the three winning teams. Finally, Section \ref{summary} summarizes the findings and discusses our results. The full dataset, as well as the benchmark consisting of the baseline models, the winning solutions, and the aggregations, are available at \href{https://gitlab.com/smarter-mobility-data-challenge/tutorials}{https://gitlab.com/smarter-mobility-data-challenge/tutorials} and distributed under the Open Database License (ODbL).

\section{EV charging dataset and target}
\label{EV_charging}

\paragraph{Dataset description}
\label{sec:dataset}

\begin{figure}
        \centering
        \includegraphics[width = \textwidth]{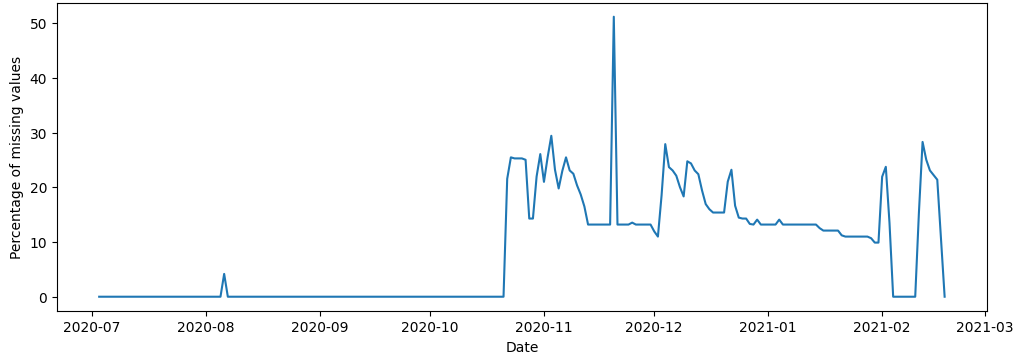}
        \caption{Percentage of missing values per day}
        \label{fig:MissingValues}
\end{figure}

The dataset is based on the real-time charging station occupancy information of the Belib network, which is available on the Paris Data platform (ODbL) \cite{paris_data}. The Belib network consists of 91 charging stations in Paris, each offering 3 plugs. Because Paris Data does not store the data it daily produces, the EDF R\&D team has set up a pipeline to collect this data every 15 minutes from July 2020 on the platform's dedicated API.  The data is then stored in a data lake based on Hadoop technologies (HDFS, PySpark, Hive, and Zeppelin). 
For the purpose of the challenge, the data has been divided into a training set and a test set. The training set contains $D_{train}$ points ranging from 2020-07-03 00:00 to 2021-02-18 23:45 CET. The test set contains $D_{test}$ points ranging from 2021-02-19 00:00 to 2021-03-10 23:45 CET. The test set has been divided into two subsets: a public set for validation purposes and a private set used to quantify the performance while minimising the risk of overestimating the performance of potential overfitting methods. To create the public and the private sets, the dataset was split into three time periods. The first is assigned to the public set, and the third to the private set. Then, we randomly draw points from the second period to assign $20\%$ of them to the public set and the rest to the private set. Another challenge related to the availability of the data is the occurrence of missing values after 2020-10-22, as illustrated by Figure \ref{fig:MissingValues}. 

\begin{figure}
    \centering
    \includegraphics[width=\linewidth]{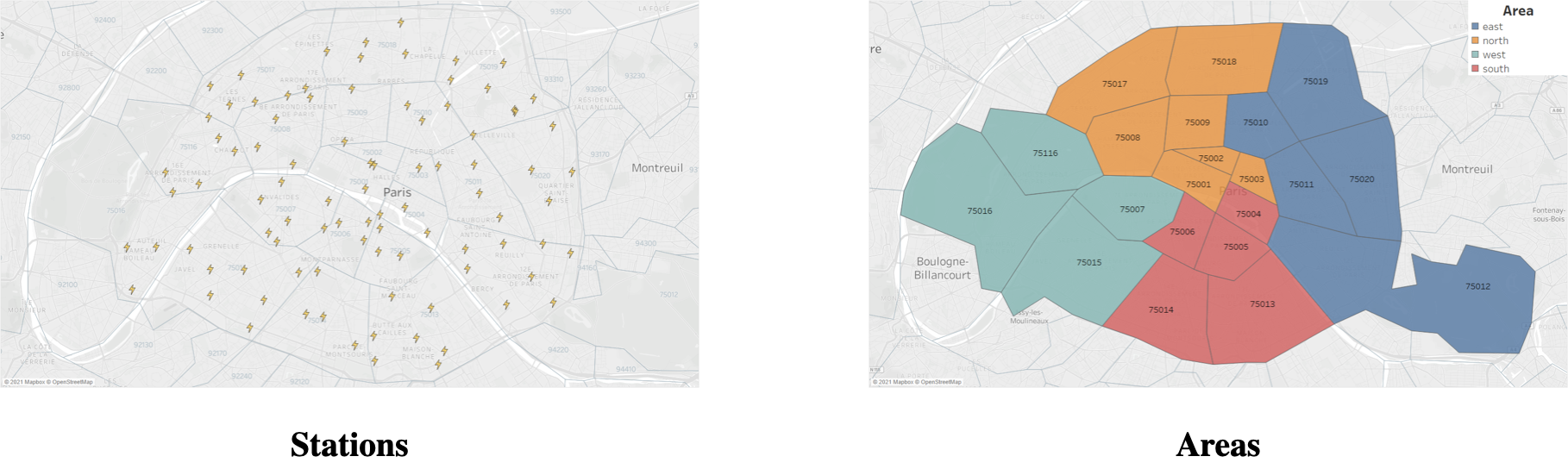}
    \caption{The 91 stations (yellow dots on the left) and the 4 areas of Paris (colored on the right)}
    \label{areas_stations}
\end{figure}

At any given time, a plug takes one of four states. Either it is plugged into a car and provides electricity, corresponding to state $c$ (charging). Either it is connected to a car that is already fully charged, corresponding to state $p$ (passive). Either the plug is free, corresponding to state $a$ (available). The special state $o$ (other) regroups the cases where the plug does not work. We denote by $y_{t,k} = (a_{t,k} , c_{t,k} , p_{t,k} , o_{t,k}) \in \{0,1,2,3\}^4$ the vector representing the state of station $k\in \{1, \hdots, 91\}$ at time $t$, where $a_{t,k}$ is the number of available plugs, $c_{t,k}$ the number of charging plugs, $p_{t,k}$ the number of passive plugs, and $o_{t,k}$ the number of other plugs, at station $k$ and time $t$. Of course, $a_{t,k}+c_{t,k}+p_{t,k}+o_{t,k}=3$. The additional variables associated with station $k$ are
\begin{itemize}
    \item temporal data: \textit{date}, \textit{tod}, \textit{dow}, and  \textit{trend},
    \item spatial data: \textit{latitude}, \textit{longitude}, and \textit{area} (south, north, east, and west) of the station.
\end{itemize}
The \textit{dow} and \textit{tod} features return respectively the day of week (1 for Monday to 7 for Sunday) and the time of day, corresponding to the 15 minute step position in the day (0 for 00:00:00 to 95 for 23:45:00). The \textit{trend} feature returns the numerical conversion of the time index. The data is then aggregated into 4 areas of about 20 stations each, as shown in Figure \ref{areas_stations}.

\paragraph{Evaluation}
\label{eval}
The goal is to forecast the states of the different plugs at 3 hierarchical levels: individual stations (denoted by $y_{t,i}$, for $i \in \{1 \ldots 91\}$), areas ($y_{t,\text{south}}$, $y_{t,\text{north}}$, $y_{t,\text{east}}$, and $y_{t,\text{west}}$), and global ($y_{t,\text{global}}$), where $y_{t,\text{zone}} = \sum_{i \in \text{zone}} y_{t,i}$ is the sum of the plugs per state in a zone (south, north, east, west, or global). Therefore, let $\displaystyle z_t= (y_{t,1},\ldots, y_{t,91}, y_{t \text{south}},  y_{t,\text{north}}, y_{t,\text{east}}, y_{t,\text{west}},  y_{t,\text{global}})$ be the aggregated matrix containing the statutes of all stations at the different hierarchical levels at time $t$.  The goal is therefore to provide the best estimator $\hat z$ of $z$. Performance is evaluated by the the  following loss function, which encodes each hierarchical level as a penalty.
\begin{equation}
    L(z_t, \hat{z}_t) = \ell_{\text{station}}(z_t, \hat{z}_t) + \ell_{\text{area}}(z_t, \hat{z}_t) + \ell_{\text{global}}(z_t, \hat{z}_t),
    \label{eq:testLoss}
\end{equation}
where
\begin{align*}
    \ell_{\text{station}}(z_t, \hat{z}_t) &= \sum_{k=1}^{91} \|y_{t,k}- \hat{y}_{t,k}\|_1, \\
        \ell_{\text{area}}(z_t, \hat{z}_t) &= \sum_{\text{zone} \in \{ \text{south},  \text{north}, \text{east}, \text{west}\}} \|y_{t,\text{zone}}- \hat{y}_{t,\text{zone}}\|_1, \\
        \ell_{\text{global}}(z_t, \hat{z}_t) &= \|y_{t,\text{global}}- \hat{y}_{t,\text{global}}\|_1,
\end{align*}
and where $\|x\|_1 = \sum_{k=1}^{p} |x_k|$ is the usual $\ell^1$ norm on $\mathbb{R}^p$. Note that $L(z_t, \hat{z}_t) = \|z_t - \hat z_t\|_1$.

\paragraph{Baseline models}
An initial baseline of two simple models is established before evaluating the performance of more complex solutions. The first naive estimator of 
$z_t$
consists in taking the median of each coefficient of 
$z_t$ over $D_{train}$ for a given value of the temporal data. 
This amounts to computing the median per type of day and quarter-hour, i.e.,
\begin{equation*}
    \hat{z}_{t} = \underset{t' \in Cal_{t}}{\mathrm{median}}\{z_{t'}\},
\end{equation*}
where
\begin{equation*} 
    Cal_{t} = \{t' \in D_{train}, \; \textrm{dow}(t') = \textrm{dow}(t)\} \cap \{ t' \in D_{train}, \;  \textrm{tod}(t') = \textrm{tod}(t)\}.
\end{equation*}
Notice that the set $Cal_t$ corresponds to the timestamps of the same day of the week and the same hour of the day. \label{CatBoostSection}
The second baseline model is the parametric model called \href{https://CatBoost.ai/}{CatBoost}. 
It is a tree-based gradient boosting algorithm that specializes in regression for categorical data.  It is implemented using the python library \texttt{CatBoost} \cite{CatBoost} and has demonstrated excellent performance for a great variety of regression tasks \cite{daoud2019comparison, huang2019evaluation, hancock2020CatBoost}. The performance of these models is shown by the dotted lines in Figure \ref{fig:ranking}.

\section{Winning Solutions}
\label{winning}
\begin{figure}
    \centering
    \includegraphics[width=\textwidth]{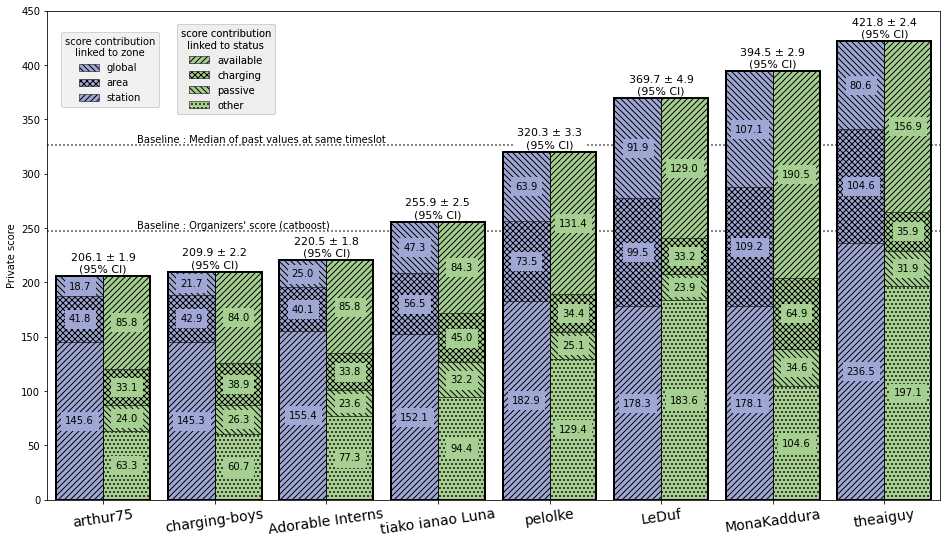}
    \caption{Ranking of the competitors}
    \label{fig:ranking}
\end{figure}

This section describes the methods used by the three winning teams. The ranking of the competitors is shown in Figure \ref{fig:ranking}, where the confidence intervals are constructed by bootstrapping. A subsection is dedicated to each of the winning teams, as their approaches are quite different and informative for the analysis of the dataset. In the last subsection, their strengths are combined thanks to aggregation methods. 

\subsection{Arthur Satouf (team Arthur75)}
\label{satouf}


\paragraph{Handling missing values} As shown in Figure \ref{fig:MissingValues}, the dataset presents a lot of missing data. Common techniques were considered to impute these \cite{pratama2016review}, including computing the mean by station, forward and backward filling, simple moving average, weighted moving average, and exponential moving weighted average (EMW) \cite{sec:ewm}. These techniques are  evaluated by measuring the mean absolute error (MAE) on a validation subset of the training set. As a result, the EMW is the most effective technique, and it is thus implemented for both forward and backward filling approaches. Specifically, we use the last 8 known values to forward fill the first 8 missing values. The same procedure is applied to backward filling.
 
\begin{figure}[htbp]
        \centering
        \includegraphics[width=\textwidth]{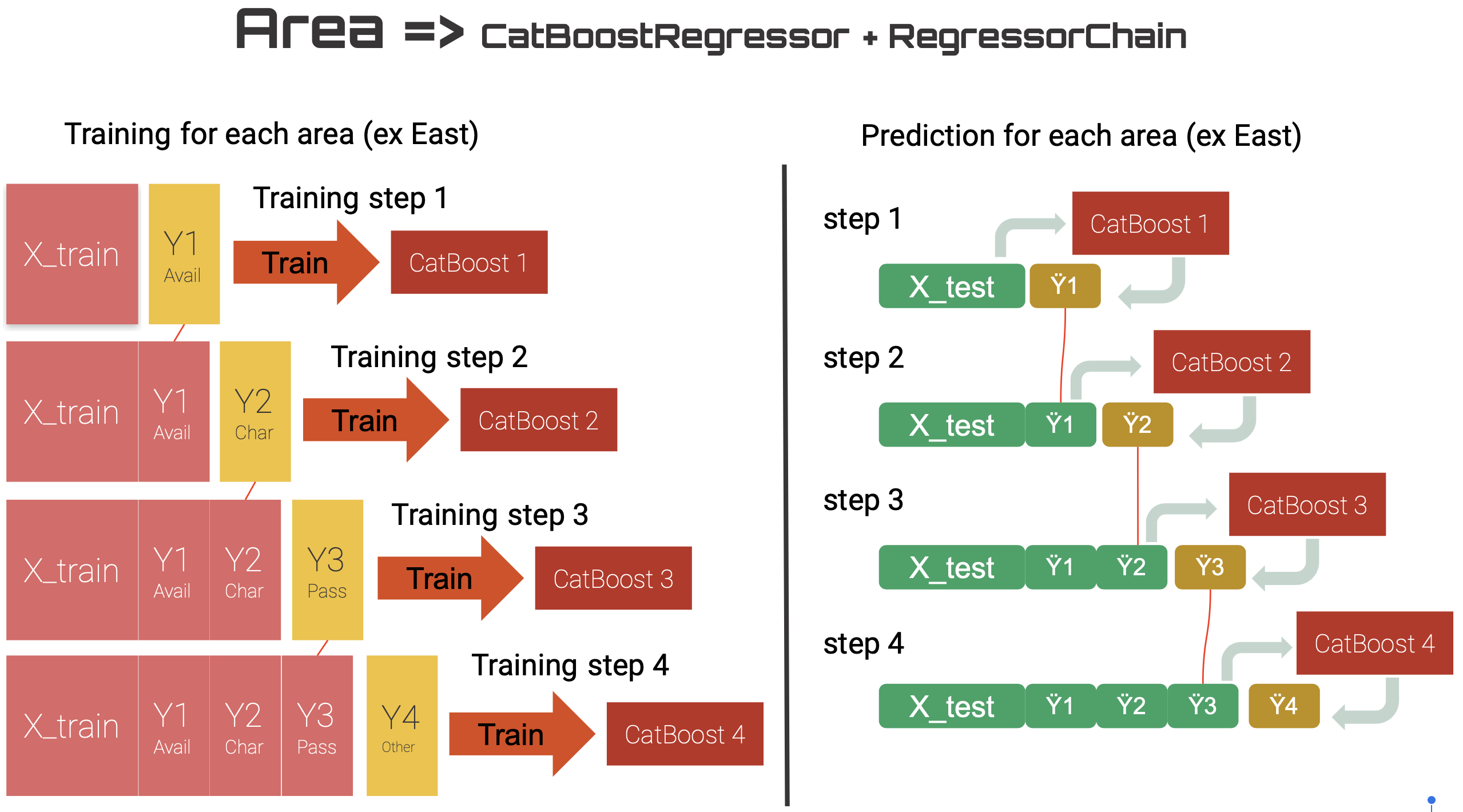}
        \caption{Using Regressor Chain with CatBoost-Regressor to train and forecast "area" data set. }
        \label{Chain}
\end{figure}

\paragraph{Model selection} 
The benchmark consists of usual time series models \cite{ahmed2010empirical, chen2016xgboost, ribeiro2020ensemble}, such as SARIMAX, LSTM, XGBoost, random forest, and CatBoost.
The evaluation metric used is the MAE, and the time series cross-validation technique is applied to evaluate the performance of the models \cite{kreiss2011bootstrap, sklearn}.  
The CatBoost algorithm is ultimately chosen for its fast optimization relying on parallelization and its ability to handle categorical data without preprocessing.
As explained in Section \ref{sec:dataset}, the states of any station $k$ satisfy at any time $t$ the equation $a_{t,k}+c_{t,k}+p_{t,k}+o_{t,k}=3$, which is enforced in the CatBoost estimator as follows.
\begin{itemize}
    \item At the station level, the problem is transformed from a multi-task regression problem to a classification problem. This is achieved by concatenating the values of each task as a string, resulting in 20 unique classes. In this approach, the sum of the four vectors always equals three, given that there are three plugs. After predicting a given target, the target is decomposed into four values. Table \ref{data_example} provides an example.

\begin{table}
\centering
\caption{Example of a data conversion to a string} 
    \begin{tabular}{ cccccc } 
    \toprule
 Given station at a given time  & Available & Charging & Passive & Other & Target \\ 
    \midrule
 14h15-16/08/2021 & 1 &2 &0 &0 & 1200 \\ 
 14h30-16/08/2021  & 0 & 1 & 1 & 1 &0111 \\ 
 14h45-16/08/2021  & 0 &0 & 3&0 &  0030 \\ 
    \bottomrule
\end{tabular}
\label{data_example}
\end{table}

    \item  At the area level, CatBoost was also used as a regression problem,  as shown in Figure \ref{Chain}. However, each area had its own model, and each area used a combination of CatBoost regressor  and Regressor-Chain \cite{src:chain}. Regressor-Chain involves building a unique model for each task and using the result of each task as an input for the next prediction model. The output of each model, along with the previous output, is then used as input for the next task. This approach helps to keep the sum of plug equal to the right number and takes into account the correlation between tasks, making the prediction more robust. 

    \item At the global level, the approach is similar to the one applied to the area level, with only 4 models as there are no longer areas. 
\end{itemize} 
A time series cross validation is used once again to tune the hyperparameters and to validate the models. It relies on the mean absolute percentage error \cite{MAPE} at the area and the global levels, and on the F-measure \cite{chen2004statistical} at the station level. In total, 21 CatBoost models are used to forecast the private datasets.

\subsection{Thomas Wedenig and Daniel Hebenstreit (team Charging-Boys)}

\paragraph{Data exploration, preprocessing and postprocessing}
Our exploratory experiments does not show any signs of a trend within the time series. Regarding stationarity, we run the Augmented Dickey–Fuller test \cite{dickey1979distribution} on the daily averages of the target values for each station and find inconclusive results. Therefore, we cannot assume stationarity for all target-station pairs, which is why we employ differencing in the construction of our ARIMA model.
As usual in statistical frameworks, we assume that the noise interferes with the high frequencies of the signal.
To denoise, we preprocess the time series by computing a rolling window average with a window size of $2.5$ hours \cite{hyndman2018forecasting}.

During our data exploration, we encounter a significant change in the behavior of the individual stations in the end of October 2020, just before the COVID-19 regulations were enforced in Paris. 
We also assume that several stations were turned off after this event, as labels were missing over large time intervals. Thus, we experiment with different methods of missing value imputation, but find that simply dropping the timestamps with missing values performs best. 
We also add custom features, namely a column indicating whether the current date is a French holiday, as well as sine and cosine transforms of \textit{tod}, \textit{dow}, the month, and the position of the day in the year.
To ensure that our regression models return integer outputs that sum to $3$ for each station and timestamp (since stations have exactly $3$ plugs), we round and rescale these predictions in a post-processing step.

\paragraph{Tree-based regression model}
Using \texttt{skforecast}, we train an autoregressive XGBoost model \cite{chen2016xgboost} with $100$ estimators. We train it on all of the 91 stations individually, each having 4 targets, resulting in $364$ models. Each model receives the last $20$ target values, as well as the sine/cosine transformed time information as input, and predicts the next target value. We also discard all features that are constant per station (e.g., station name, longitude, and latitude). The final regression model achieves a public leaderboard score of $177.67$.

\paragraph{Tree-based classification model}
To effectively enforce structure in the predictions (i.e., that they sum to $3$), we transform the regression problem discussed above into a classification problem. For a given station and timestamp, consider the set of possible target values $\mathcal{C} = \left\{ \mathbf{x} \in \{0,1,2,3\}^4 \quad \text{s.t.} \ \sum_{i=1}^4 x_i = 3 \right\}$.
We treat each element $c \in \mathcal{C}$ as a separate class and only predict class indices $\in \mathcal{I} = \{0, \dots, 19\}$ (since $|\mathcal{C}| = 20$). While $\mathcal{I}$ loses the ordinal information present in $\mathcal{C}$, this approach empirically shows competitive performance. When training a single XGBoost classifier with 300 estimators for all stations, we achieve a public leaderboard score of $178.9$. We also experiment with autoregressive classification (i.e.,including predictions of previous timestamps), but find no improvement in the validation error.

\paragraph{ARIMA model}
We fit a non-seasonal autoregressive integrated moving average (ARIMA) model \cite{box2015time} for each target-station combination.
To predict the value of a given target, we only consider the last $p = 2$ past values of the same target (in the preprocessed time series) and do not use any exogenous variables for prediction (e.g., time information).
We apply first-order differencing to the time series ($d=1$) and design the moving average part of the model to be of first-order ($q=1$).
We observe that the forecasts using these models have very low variance, i.e., each model outputs an approximately constant time series.
These predictions achieve a competitive score on the public leaderboard (third place).

\paragraph{Ensemble method}
The final model is an ensemble of the tree-based regression model, the tree-based classification model, and the ARIMA model. For a single target, we compute the weighted average of the individual model predictions (per timestamp). The ensemble weights are chosen to be roughly proportional to the public leaderboard score ($w_{\text{reg}} = 0.35$, $w_{\text{class}} = 0.25$,  $w_{\text{ARIMA}} = 0.4$).
Since the predictions of the tree-based models have high variance, we can interpret mixing in the ARIMA model's predictions as a regularizer, which decreases the variance of the final model.
As the tree-based models also use time information for their predictions, we use the entirety of the available features.

\subsection{Nathan Doumèche and Alexis Thomas (team Adorable Interns)}
\label{teamai}
\begin{figure}
        \centering
        \includegraphics[width=\textwidth]{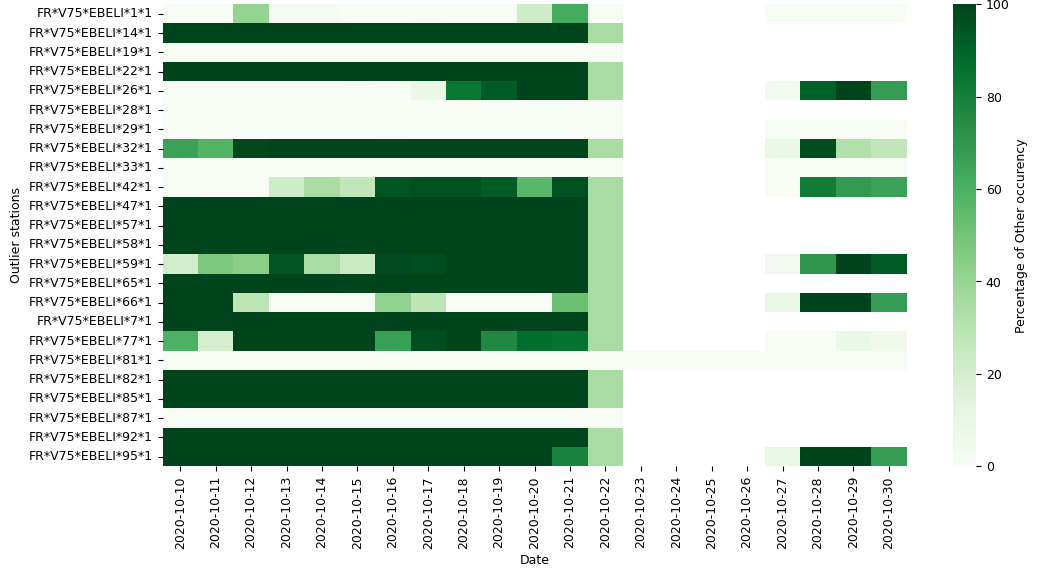}
        \caption{Percentage of state $o$ occurrences per outlier per day around 2020-10-22}
        \label{fig:Other2}
\end{figure}

\paragraph{Data analysis} Several challenges arise from the data, as shown in Figure \ref{fig:MissingValues}. 
An interesting phenomenon is the emergence of a change in the data distribution on 2020-10-22, characterized by the appearance of missing data.  
A reasonable explanation is that the detection of missing values is due to an update in the software that communicates with the stations.
The update would have taken place on 2020-10-22, allowing the software to detect new situations in which stations were malfunctioning. 
This hypothesis is supported by the fact that the stations with missing values are those that were stuck in states corresponding to the absence of a car, i.e., either the state $a$  or the state $o$ (see Figure \ref{fig:Other2}). In fact, 88\% of the stations that were stuck in either $a$  or $o$  for the entire week before 2020-10-22 had missing values on  2020-10-22.
Perhaps the users avoided the malfunctioning stations, or perhaps the users tried to connect to the station, but the plug was unresponsive, so the users went undetected. 
An important implication of this hypothesis is that the data before the change  should not be invalidated, since the behaviour of the well-functioning stations did not change. 
Another challenge of the dataset was its shortness. In fact, we expect a yearly seasonal effect due to holidays \citep{Xing2019charging} that cannot be distinguished from a potential trend because there is less than one year of data. 
All these observations suggest giving more weight to the most recent data.

\paragraph{Empirical loss} As usual in the supervised learning setting, we need to choose a model $\mathcal{F}$ to construct the estimator $\hat{z}_t \in \mathcal{F}$. 
To estimate the entire $D_{test}$ period at once, we cannot rely on online models such as autoregressive models or hidden-state neural networks (RNN, LTSM, transformers...), although they perform well for time series forecasting \citep{bryan2021survey}, and in particular for EV charging station occupancy forecasts \cite{ma2022multistep, mohammad2023energy}.
Once a model $\mathcal{F}$ is chosen, we define an empirical loss $L$ on the training data. Then, a learning procedure, such as a gradient descent, fits the estimator $\hat{z}$ that minimizes $L$, with the hope that $\hat{z}$ will minimize the expectation of the test loss \eqref{eq:testLoss} \citep{vapnik1991principles, hastie2017elements}. Given a training set $T_{train} \subseteq D_{train}$, we consider two empirical losses. The first one is $L_{equal}(\hat{z}) = \sum_{t \in T_{train}} 
\|z_t - \hat{z}_t\|_1$. It gives equal weight to all data points. The second one is $L_{exp}(\hat{z}) = \sum_{t \in T_{train}} \exp((t-t_{max})/\tau)\|z_t - \hat{z}_t\|_1$,
where $\tau = 30$ days and $t_{max} =$ 2021-02-19 00:00:00. This time-adjusted loss function is common for non-stationary processes \cite{ditzler2015learning} because it gives more weight to the most recent observations. This makes it possible to give more credit to the data after the change in the data distribution and to capture the latest effect of the trend, while using as much data as possible.

\paragraph{Benchmark phase} To compare the performance of the models, we define a training period $T_{train}$, which covers the first $95\%$ of $D_{train}$, and a validation period $T_{val}$, which covers the last $5\%$. Models are trained on $T_{train}$ to minimize $L_{equal}$ or $L_{exp}$, and then their performance is evaluated on $T_{val}$ by $L_{val}(\hat{z}) = \sum_{t \in T_{val}} \|z_t-\hat{z}_t\|_1$.  The $Mean$ model estimates $\hat{y}_{t, k}$, $\hat{A}_{t, k}$ and $\hat{G}_t$ by their mean over the training period for each value of $(tod, dow)$. Idem for the $Median$ model. They are robust to missing values since the malfunctioning of a station $k$ only affects $\hat{y}_{t,k}$. 
We compare them with the CatBoost model presented in Section \ref{CatBoostSection}. Let $C(d, i)$ be the CatBoost model of depth $d$ trained with $i$ iterations using $L_{equal}$, and $C_{exp}(d, i)$ the same model trained using $L_{exp}$. 
In this setting, we train twelve CatBoost models: one for each pair of state ($a, c, p, o$) and hierarchical level. 
After hyperparameter tuning, we found $C(4, 150)$ and $C_{exp}(5,200)$ to be the best models in terms of tradeoff between performance and number of parameters, knowing that early stopping and a small number of parameters prevent overfitting \citep[see, e.g.,][]{ying2019An}. All of these models take advantage of the fact that malfunctioning stations tend to stay in specific states.

\paragraph{Validation phase} The contest organizers allowed participants to test their models on a subset $T_{val}$ of $D_{test}$. In this phase, we trained  our best models on the entire $D_{data}$ period and sent them to be tested with the test loss \eqref{eq:testLoss}. Table \ref{fig:perf} shows that the hierarchy of the models is preserved.
\begin{table}
    \centering
    \caption{Evaluation of the performance of the Adorable Interns' models  in both phases}
    \begin{tabular}{ccccc}
        \toprule & Mean & Median &$ C(4, 150)$ & $ C_{exp}(5, 200)$\\
        \midrule
         Benchmark Phase & 316 & 309 & 292 & \bf{261}\\
         Validation Phase& 323 & 303 & 233 & \bf{189}\\
            \bottomrule
    \end{tabular}
    \label{fig:perf}
\end{table}
The submitted model  was therefore $C_{exp}(5, 200)$. Note that this model is also interesting because its small number of parameters ensures robustness and scalability. In addition, tree-based models are quite interpretable, which is paramount for operational use \cite{jabeur2021CatBoost}.

\subsection{Aggregation of the teams' forecasts}

Naive aggregations of uncorrelated estimators are known to have good asymptotic \citep{tsybakov2003optimal} and online \cite{cesa2006prediction} properties. In practice, they often achieve better performance than the individual estimators \citep[see, e.g.,][]{bojer2021kaggle, mcandrew2021aggregating}. 
Once again, this is the case  in this challenge, as evidenced by 
Table \ref{table_score_target_agg2}, which shows the performance of the top 3 teams compared to two aggregation techniques. The uniform aggregation --denoted by \textit{Uniform agg.}-- corresponds to the mean of each team's prediction, while the weighted aggregation --denoted by \textit{Weighted agg.}-- is computed by gradient descent using the MLpol algorithm \cite{gaillard2014second} to minimise the error on the training set. Notice how the weighted aggregation outperforms the other forecasts for the total loss, as well as for all the subdivisions of the loss by hierarchical level and by state.

\begin{table}[ht]
\centering
\caption{Score by target of the top 3 teams and aggregations} 
\begin{tabular}{rrrrrrrrr}
  \toprule
 & Available & Charging & Passive & Other& Station & Area & Global & Total \\ 
  \midrule
Arthur75 & 86 & \textbf{33} & 24 & 63 &  145 & 42 & \textbf{19} & 206 \\ 
  Charging Boys & 84 & 39 & 26 & 61 & 145 & 43 & 22 & 210 \\ 
  Adorable Interns & 86 & 34 & 24 & 77 & 155 & \textbf{40} & 25 & 220  \\ 
  Uniform agg. & 83 & \textbf{33} & \textbf{22} & 63 &  142 & \textbf{40} & 20 & 202 \\ 
  Weighted agg. & \textbf{82} & \textbf{33} & \textbf{22} & \textbf{59} & \textbf{137} & \textbf{40} & \textbf{19} & \textbf{196}  \\ 
   \bottomrule
\end{tabular}
\label{table_score_target_agg2}
\end{table}

\section{Summary of findings and discussion}
\label{summary}

This paper presents a dataset in the context of hierarchical time series forecasting of EV charging station occupancy, providing valuable insights for energy providers and EV users alike. The results discussed in this paper focus on two key aspects of the Smarter Mobility Data Challenge: data preprocessing and training the hierarchical forecasting model. As with any real-world data set, specific techniques are required to deal with missing data and outliers  (see, e.g., Section \ref{satouf}). Data preprocessing was a crucial step, and the addition of relevant exogenous features, such as the national holidays calendar, significantly improved the results, in contrast to the use of a more complex model that could not perform well without prior augmentation of the data samples.  All three winning solutions described in this paper were robust enough to maintain a high private test score, showing good generalization of the models.  The choice of the empirical cost function that drives the training process produced the best results when more recent data points were given greater weight (see, e.g., Section \ref{teamai}).  Aggregating the forecasts of the three winning teams even yielded a better global score, with a notable improvement at the station level. The hierarchical models presented are promising and could help improve the overall EV charging network. 
This open dataset is interesting because it encompasses many real-world problems related to time series matters, such as missing values, non-stationarities, and spatiotemporal correlations. In addition, we believe that sharing the benchmark models derived from this challenge will be useful for making comparisons in future research.

\paragraph{Limitations} The deployment of electric vehicles (EVs) is progressing at a remarkable pace \cite{sathiyan2022comprehensive}, making any dataset merely a snapshot of a swiftly evolving world \citep[see also][]{hecht2021predicting}. 
To enhance forecasting accuracy, additional features could be incorporated into a dataset. Numerous covariates, such as mobility and traffic information, meteorological data, and vehicle characteristics, could be included. In a forthcoming release of the dataset, we intend to incorporate traffic data and meteorological data. 

\paragraph{Ethical concerns} To the best of our knowledge, our work does not pose any risk of security threats or human rights violations. Knowing when and where someone plugs in their EV could lead to a risk of surveillance.
However, this dataset does not contain any personal information about the user of the plug or their car, so there is no risk of consent or privacy.

\section*{Acknowledgments}
We thank Cédric Villani, Jean-Michel Poggi, and Marc Schoenauer for being part of the jury and for their insightful comments on the algorithms and on the paper. We thank Jerome Naciri, and Jean-Yves Moise for their help in organizing the challenge. Moreover, we thank all the contestants for their original solutions. Finally, the authors thankfully acknowledge the \textit{Manifeste IA} network of French industrial companies and the TAILOR European project on trustworthy AI for founding this challenge. 

\bibliography{references.bib}

\newpage

\section*{Checklist}


\begin{enumerate}

\item For all authors...
\begin{enumerate}
  \item Do the main claims made in the abstract and introduction accurately reflect the paper's contributions and scope?
    \answerYes{See Section \ref{intro}.}
  \item Did you describe the limitations of your work?
    \answerYes{See Section \ref{summary}, paragraph \textbf{Limitations}}.
  \item Did you discuss any potential negative societal impacts of your work?
    \answerYes{See Section \ref{summary}, paragraph \textbf{Ethical Concerns}.}
  \item Have you read the ethics review guidelines and ensured that your paper conforms to them?
     \answerYes{}
\end{enumerate}

\item If you are including theoretical results...
\begin{enumerate}
  \item Did you state the full set of assumptions of all theoretical results?
    \answerNA{}
	\item Did you include complete proofs of all theoretical results?
    \answerNA{}
\end{enumerate}

\item If you ran experiments (e.g. for benchmarks)...
\begin{enumerate}
  \item Did you include the code, data, and instructions needed to reproduce the main experimental results (either in the supplemental material or as a URL)?
    \answerYes{See the end of  the Abstract, or the end of the introduction to get the link to the GitLab account with the code, data, and instructions (as a Readme).}
  \item Did you specify all the training details (e.g., data splits, hyperparameters, how they were chosen)?
    \answerYes{The global description of the algorithms is detailed throughout the paper. The specific choice of parameters is detailed in the code.}
	\item Did you report error bars (e.g., with respect to the random seed after running experiments multiple times)?
    \answerYes{See Figure \ref{fig:ranking}.}
	\item Did you include the total amount of compute and the type of resources used (e.g., type of GPUs, internal cluster, or cloud provider)?
    \answerYes{The data set is relatively small (about 180 Mb) and the algorithms of the benchmark are not computationally intensive. The experiments can be re-run on CPU.}
\end{enumerate}

\item If you are using existing assets (e.g., code, data, models) or curating/releasing new assets...
\begin{enumerate}
  \item If your work uses existing assets, did you cite the creators?
    \answerYes{See Section \ref{EV_charging}, paragraph \textbf{Dataset description}.}
  \item Did you mention the license of the assets?
    \answerYes{See Section \ref{EV_charging}, paragraph \textbf{Dataset description}.}
  \item Did you include any new assets either in the supplemental material or as a URL?
    \answerNA{}
  \item Did you discuss whether and how consent was obtained from people whose data you're using/curating?
    \answerYes{See Section \ref{summary}, paragraph \textbf{Ethical concerns}}
  \item Did you discuss whether the data you are using/curating contains personally identifiable information or offensive content?
    \answerYes{See Section \ref{summary}, paragraph \textbf{Ethical concerns}}.
\end{enumerate}

\item If you used crowdsourcing or conducted research with human subjects...
\begin{enumerate}
  \item Did you include the full text of instructions given to participants and screenshots, if applicable?
    \answerNA{}
  \item Did you describe any potential participant risks, with links to Institutional Review Board (IRB) approvals, if applicable?
    \answerNA{}
  \item Did you include the estimated hourly wage paid to participants and the total amount spent on participant compensation?
    \answerNA{}
\end{enumerate}

\end{enumerate}


\end{document}